# A Method of EV Detour-to-Recharge Behavior Modeling and Charging Station Deployment

Tianshu Ouyang[†, 1], Jiahong Cai[†, 2], Yuxuan Gao[†, 3], Xinyan He[†, 4], Huimiao Chen[†, 1, *], Kexin Hang[1]

[†] These authors contributed equally to this work.

[1] Sparkzone Institute; [2] University of Chicago; [3] The Wharton School of the University of Pennsylvania; [4] Princeton University

[*] hchen@sparkzone.org

***Abstract*—Electric vehicles (EVs) are increasingly used in transportation. Worldwide use of EVs, for their limited battery capacity, calls for effective planning of EVs charging stations to enhance the efficiency of using EVs. This paper provides a methodology of describing EV detouring behavior for recharging, and based on this, we adopt the extra driving length caused by detouring and the length of uncompleted route as the indicators of evaluating an EV charging station deployment plan. In this way, we can simulate EV behavior based on travel data (demand). Then, a genetic algorithm (GA) based EV charging station sitting optimization method is developed to obtain an effective plan. A detailed case study based on a 100-node 203-branch transportation network within a 30 km × 30 km region is included to test the effectiveness of our method. Insights from our method may be applicable for charging station planning in various transportation networks.**

## I. Introduction

The continued growth in use of conventional vehicles in transportation will unavoidably cause natural gasoline resource dwindling. As an unrenewable fuel source, gasoline will be in shortage in the near future due to human's immoderate exploitation, which will also result in severe environmental degradation, e.g., excessive emission of greenhouse gases. With the awareness of the limited fossil fuels, governments have implemented numerous policies to promote purchase and use of electric vehicles (EVs), which rely on electricity that can be generated from wind power, hydraulic power, etc. However, the peculiarities of EVs, including insufficient battery capacity to reach a remote destination, commonly referred as "range anxiety", and frequent demand to charge, have been the dominant hinderances to market acceptance of EVs. Therefore, EV users are restricted to drive only short trips and less annual miles due to inconvenience to charge, which not only discourages customer adoption of EVs but also impedes the transition from gasoline-powered vehicles to the more environmentally friendly EVs.

Therefore, effective EV charging stations planning is urgently needed to counter the issues of inaccessibility to charging stations and excessive increment of driving due to recharging, or else EVs will fail to receive public market acceptance. Incorporating EV charging station networks into current transportation networks requires thorough considerations and simulations because of the complicacies of urban areas. However, with the rapidly growing population density and increasing demand for vehicles, effective planning of charging stations in city areas can bring sizable social benefits.

In existing literature, there are studies focusing on planning of different types of energy supply infrastructure in transportation, e.g., gas stations and charging stations. Reference [1] explores the choosing of gas stations and the factors that influence the decision. In [2], the authors introduce a model for positioning the optimum location of a small size fuel station with the help of Geospatial Information System, and methods such as Boolean logic, index overlay, and fuzzy operators are demonstrated. Sweda *et al.* employ an agent-based decision support system presented for identifying patterns in residential EV ownership and driving activities to enable strategic deployment of new charging infrastructure [3]. The authors of [4] present a multi-objective collaborative planning strategy to deal with the optimal planning issue in integrated power distribution and EV charging systems. The user equilibrium-based traffic assignment model is integrated to address the maximal traffic flow capturing problem, and decomposition based multi-objective evolutionary algorithm is employed to seek the non-dominated solutions. Reference [5] proposes a two-step model for allocating EV recharging stations that first quantifies the road information into data points and subsequently converges into 'demand clusters' over an urbanized area by hierarchical clustering analysis; optimization techniques are then applied on the demand clusters with the aim of meeting the supplies and demands, and certain constraints and cost factors are considered. Dong *et al.* investigate the influence of public charging stations deployment on increasing EV market penetration; an activity-based assessment method serves to test the feasibility for the heterogeneous traveling population, and genetic algorithm (GA) is used to find the optimal locations for charging stations; Dong *et al.* base on GPS-based travel survey data to conduct case studies [6]. Payam *et al.* propose a dynamic approach to the optimization of charging station locations and sizing that best suffice the charging demands and minimize power loss in the transportation of electricity from central power grids [7]. The research of Xiong *et al.* focuses on the charging behaviors and mutual impact on varied traffic conditions of EV drivers [8]. Li *et al.* [9] establish mathematical model based on bi-level programming to formulate government's infrastructure location strategy, the company's fleet composition and routing plan in order to promote EV adoption in companies. A hybrid heuristic method is then employed to solve the formulated problems. Bouguerra *et al.* [10] study the real case of the center of Tunis City, Tunisia with existing data to find optimized location and size of charging stations of this specific area. Reference [10] uses the case in the central urban area, a 30 km by 30 km area, in Beijing to study the data-driven planning of electric taxis considering the optimization of both investment and operation costs.

Despite the already done work, describing EV driving behavior and evaluating an EV charging station sitting plan are still challengeable due to factors including limited data of EVs. How to model the EV behavior from conventional vehicle route data and use it in station planning is a meaningful problem which has not been investigated adequately. Hence, in this work, we provide an EV behavior model and a station planning method derived from it. The main procedures and contributions of the paper are summarized below.

- Present a methodology of describing the EV detouring behavior for recharging.
- Adopt the extra driving length caused by detouring and the length of uncompleted route as the indicator to evaluate of station planning.
- Develop a GA based method to plan EV charging station in a transportation network with travel data not from EVs.
- Study a detailed case with a 100-node 203-branch transportation network within a 30 km × 30 km region to test the effectiveness of the planning.

The remainder of the paper is organized as follows. Section II illustrates the methodology of describing EV detouring behavior for recharging. Section III introduces the GA for EV charging station planning and Section IV shows the case. Section V concludes.

## II. Methodology of Describing EV Detouring Behavior for Recharging

In this section, we introduce our methodology of describing the EV detouring behavior for recharging. We consider a transportation network consisting of a series of nodes (intersections) and branches (roads) with vehicle travel data given. The network here is limited to a connected graph because an unconnected graph can be treated as multiple connected graphs. Vehicle travel data is in form of a set of routes, each of which includes a series of nodes that the vehicle needs to pass in order. Besides, for each route record, there is a corresponding initial state of charge (SOC).

Detouring to recharge their batteries is the key difference between EV and conventional vehicle travel behavior. Herein, we first introduce some assumptions for the description of EV detouring. 1) An EV only charges once in a detour, but it can detour multiple times in a route deviating from different branches. This consideration is for the purpose of making the detour rule simple, clear, reasonable and practical. 2) The length of each branch is the shortest length between the two connected nodes. This assumption ensures that taking detour will absolutely make the length of driving longer than not taking detour, so the times of taking detour is intended to be minimized. 3) Every node in a route must be passed to complete the route. If an EV detours from a node to charge, it must return to the original route by passing the next node after charging. In other words, all nodes in a route must be passed.

Before elaborating on the detour rules, we define some notations. We use $L_{i-j}$ ($L_{i-j-k}$) to denote the length of the shortest path from node *i* to node *j* (from node *i* to node *j* to node *k*) and $L_{detour_n}$ to denote the difference of the lengths of the detour path and the branch, i.e., $L_{detour_n}=L_{i-c-j}-L_{i-j}$ (*c* is a node with charging station, *n* represents that node *i* is the *n*th node in the route), and use $L_{rest}$ to denote the length of the uncompleted part of the route.

The detour rules include multiple steps of judging, computations, and recording. When an EV arrives at a node of a route, it needs to conduct the following steps to determine whether and how to detour. For a route *r* whose nodes are $\boldsymbol{N}_r=\{x_1,x_2,x_3,\cdots,x_N\}$, if the remained SOC of the EV at $x_n$ can fulfill the driving from $x_n$ to $x_{n+1}$, we let $L_{detour_n}=0$. If the remained EV SOC cannot fulfill the driving from $x_n$ to $x_{n+1}$, we select all charging stations that can be reached with the remained SOC as a set $\boldsymbol{C}_{r,x_n}=\{c_1,c_2,c_3,\cdots\}$. If $\boldsymbol{C}_{r,x_n}$ is an empty set, i.e., no charging station is accessible, the EV will not be able to detour to charge. We calculate $L_{rest}$ and record a new $L=L_{rest}+\sum_{i=1}^{n-1}L_{detour_i}$. If $\boldsymbol{C}_{r,x_n}$ is not empty, there are charging stations accessible for the EV. If the EV does not choose to detour to charge, we calculate $L_{rest}$ and record a new $L=L_{rest}+\sum_{i=1}^{n-1}L_{detour_i}$. If EV detours to charge, we calculate $L_{x_n-c-x_{n+1}}$ for all $c\in\boldsymbol{C}_{r,x_n}$ and select the station with the minimum value for this charging, and calculate $L_{detour_n}$. If there exist multiple stations with the same minimum $L_{x_n-c-x_{n+1}}$, which produces a set $\boldsymbol{C}_{\min}$, we select the station with the minimum $L_{c-x_{n+1}}$ for this charging and calculate $L_{detour_n}$ because in this way the EV will have a higher SOC when arriving at the next node. If there exist more than one stations with both the same minimum $L_{x_n-c-x_{n+1}}$ and $L_{c-x_{n+1}}$, we select one station randomly from these charging stations and calculate $L_{detour_n}$. Fig. 1 gives an example of how to select the station for charging when the SOC is insufficient, where the green nodes within the yellow dash circle, i.e., $c_1$, $c_2$, $c_3$, $c_4$ and $c_5$, are with accessible charging stations, while the station at $d_1$ is inaccessible. The green nodes on the blue dashed ellipse, i.e., $c_1$, $c_2$ and $c_3$, have the smallest $L_{x_n-c-x_{n+1}}$, and among them $c_2$ and $c_3$ have the smallest $L_{c-x_{n+1}}$. Thus, $c_2$ and $c_3$ are both the optimal locations for the charging, and the EV will randomly detour to charge at $c_2$ or $c_3$.

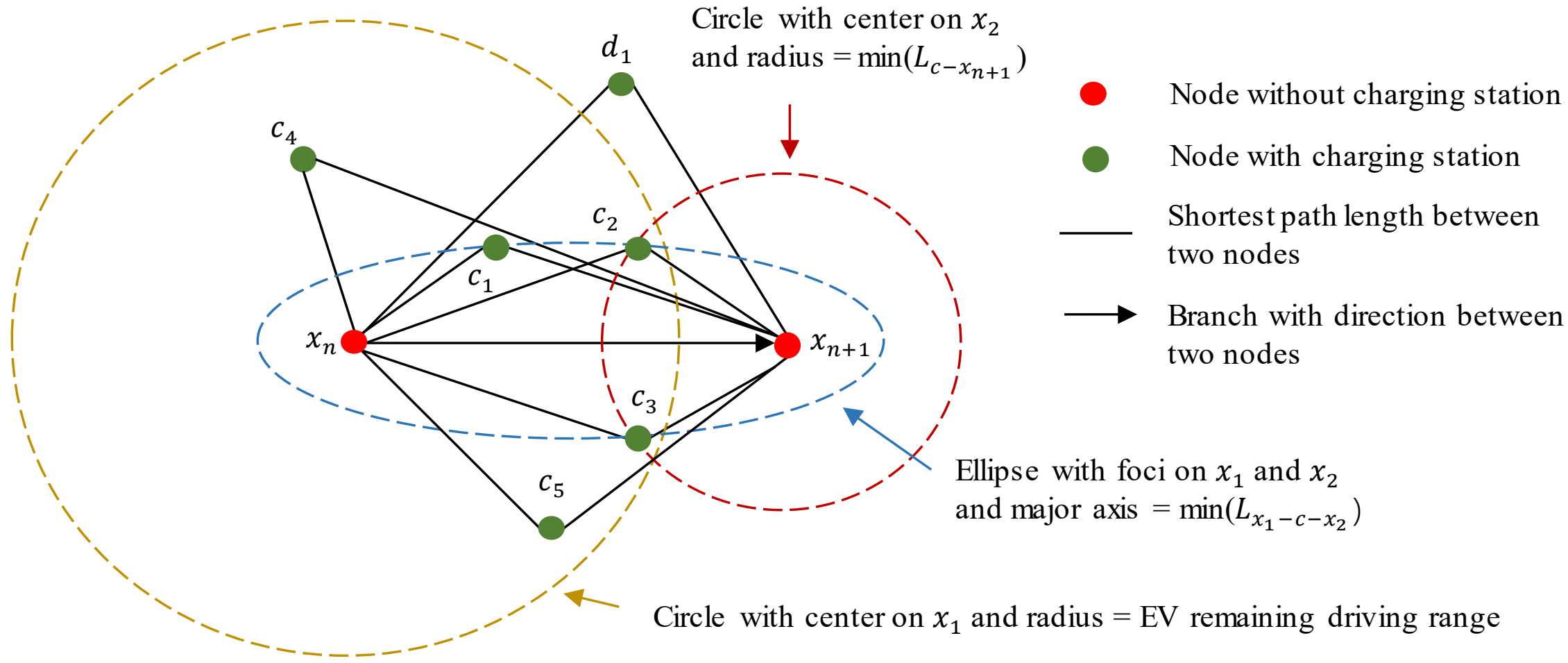

Fig. 1. Diagram of describing judging and computations in EV detouring behavior.

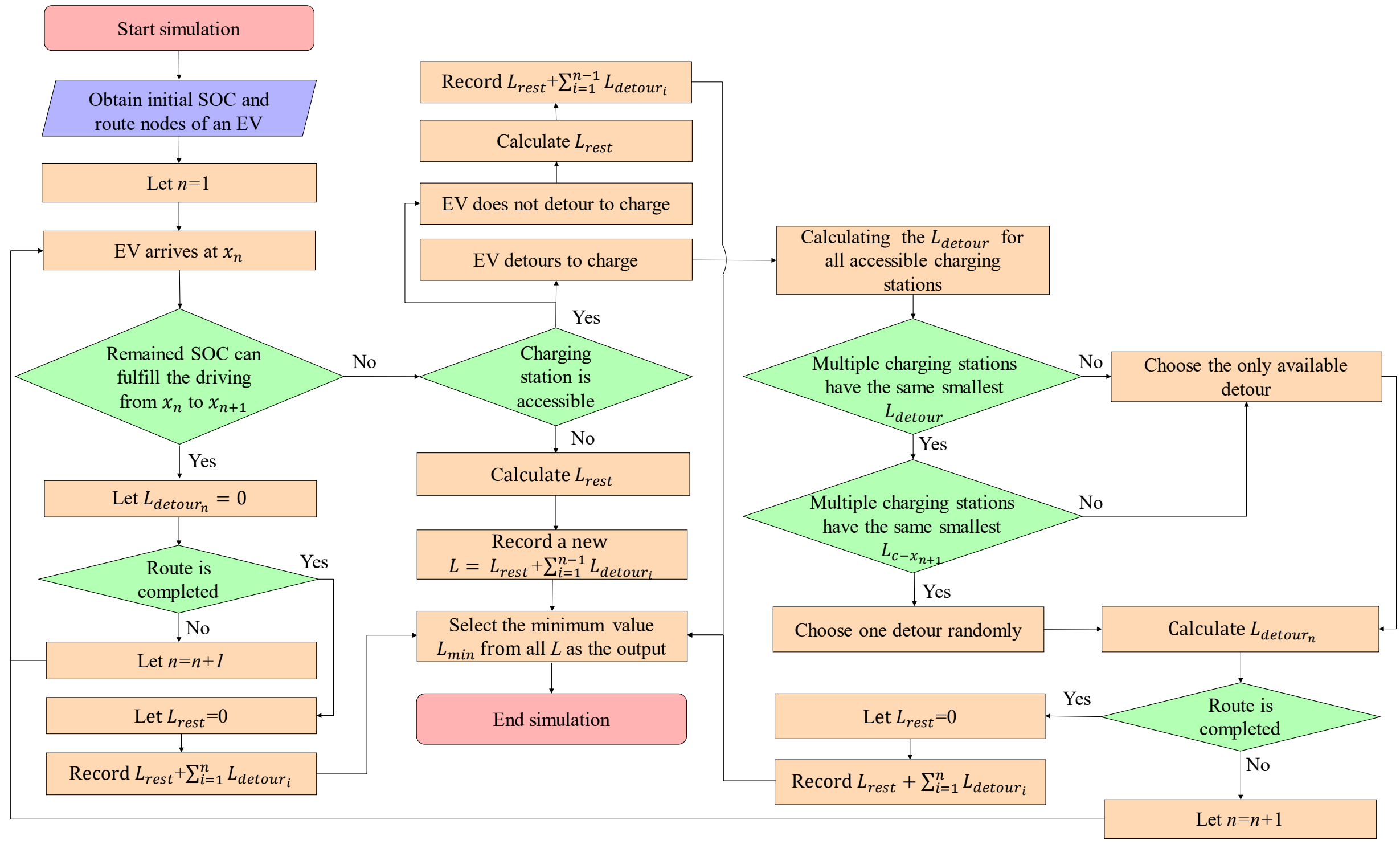

Fig.2. Flow chart of describing EV detouring behavior for recharging.

After driving from $x_n$ to $x_{n+1}$, if the route is completed, i.e., the current node is $x_N$, we let $L_{rest}=0$, calculate $\sum_{i=1}^{n} L_{detour_i}$, the total increment in driving length due to detouring, and record a new $L = L_{rest} + \sum_{i=1}^{n} L_{detour_i}$. If the route is not completed, we let $n=n+1$ and continue the simulation by repeating the above detour steps. At last, we select the minimum value $L_{\min}$ from all $L$ as the output and end simulation.

The above description of EV detouring behavior is illustrated by Fig. 2.

## III. Genetic Algorithm for Charging Station Siting Optimization in Transportation Networks

GA is adopted and utilized to obtain an optimal charging station deployment plan in given transportation networks with EV travel data. In this section, we describe our designs of key steps, i.e., initialization, selection, crossover and mutation, of the GA.

### A. Initialization

The initial population is randomly generated, where a chromosome of an individual consists of a given number of the nodes that are selected from all candidate nodes as charging station locations. All individuals form a set of charging station deployment plans, aka population in the GA. Each plan is unique in the set, and the serial numbers of nodes are sorted in ascending order in an individual's chromosome, mathematically expressed by a vector. None of the chosen nodes have a duplicate in the chromosome. The size of each chromosome equals the predetermined number of new stations.

### B. Selection

Selection is the part where the new generation is selected from the old. In this process, the roulette choosing method is used. The probability of being chosen for each chromosome is determined by their fit values. In our process, a smaller (better) fit value brings a higher probability for the individual to be selected as a member of a new generation, which may have multiple same individuals.

### C. Crossover

During the process of the GA crossovers, a predetermined crossover probability restricts the occurrence of a crossover. If a crossover happens, two individuals' chromosomes are randomly selected to perform the operation. The mutual and exclusive part of two chromosomes are collected into two separate parts. The elements in the exclusive part are randomized and split into two equally long parts, which later each combine with the one set of mutual elements to form two new chromosomes. These new chromosomes rewrite the parent chromosomes in the population for the purpose of maintaining population size. Fig. 3 gives an example.

### D. Mutation

The mutation process changes an individual's chromosome, in which similar to the crossover the occurrence is controlled by a limiting probability. If an individual is requested to mutate, a random node in its chromosome will be replaced by a new one randomly selected from the remaining node candidates. The population size is maintained here. Fig. 4 gives an example.

## IV. Case Studies

### A. Transportation Network Settings and Travel Demand Generation

In this subsection, an assumed transportation network with 100 nodes and 203 branches is constructed to simulate a real-world transportation network and verify the methodology proposed in the previous text. The generation of graph obeys the following conditions: 1) any two branches do not intersect except at nodes, which avoids additional nodes; 2) the nodes and branches form a connected graph, which ensures the connectivity of the transportation grid.

The network projects to a region sizing 30 km × 30 km. Each of the nodes is assigned an identity from four types of functionalities, i.e., commercial areas, industrial areas, residential areas, and other places. As shown in Fig. 6, 100 nodes are evenly distributed into the four categories with nodes of each type closely gather in cluster. All the 100 nodes are regarded as potential charging station positions. The lengths of all the branches are designed to be smaller than 7 km, which better resembles the roads in urban areas.

Around 10000 travel routes (vehicle travel demand not of EVs) are generated with their origins distributing in residential, commercial and industrial areas and other places with a probability of 45%, 30%, 15%, and 10%, respectively. A route generally includes 2 to 24 nodes. One node only exists once in a route, ensuring the resemblance of real-world travel demand. The route distribution in general refers to the analysis of traveling data in May 2016 Beijing. See Appendix for route generation method.

According to typical EVs on the market, the capacity of EV battery is set as 50 kWh and a consuming speed of 0.25 kWh/km. A corresponding initial State of Charge (SOC) of an EV is assigned to each route. The value is random between the SOC required to complete the route (EV with no charging demand is useless to the simulation since a detour is not necessary) and the minimum SOC that ensures the EV could reach the nearest charging station.

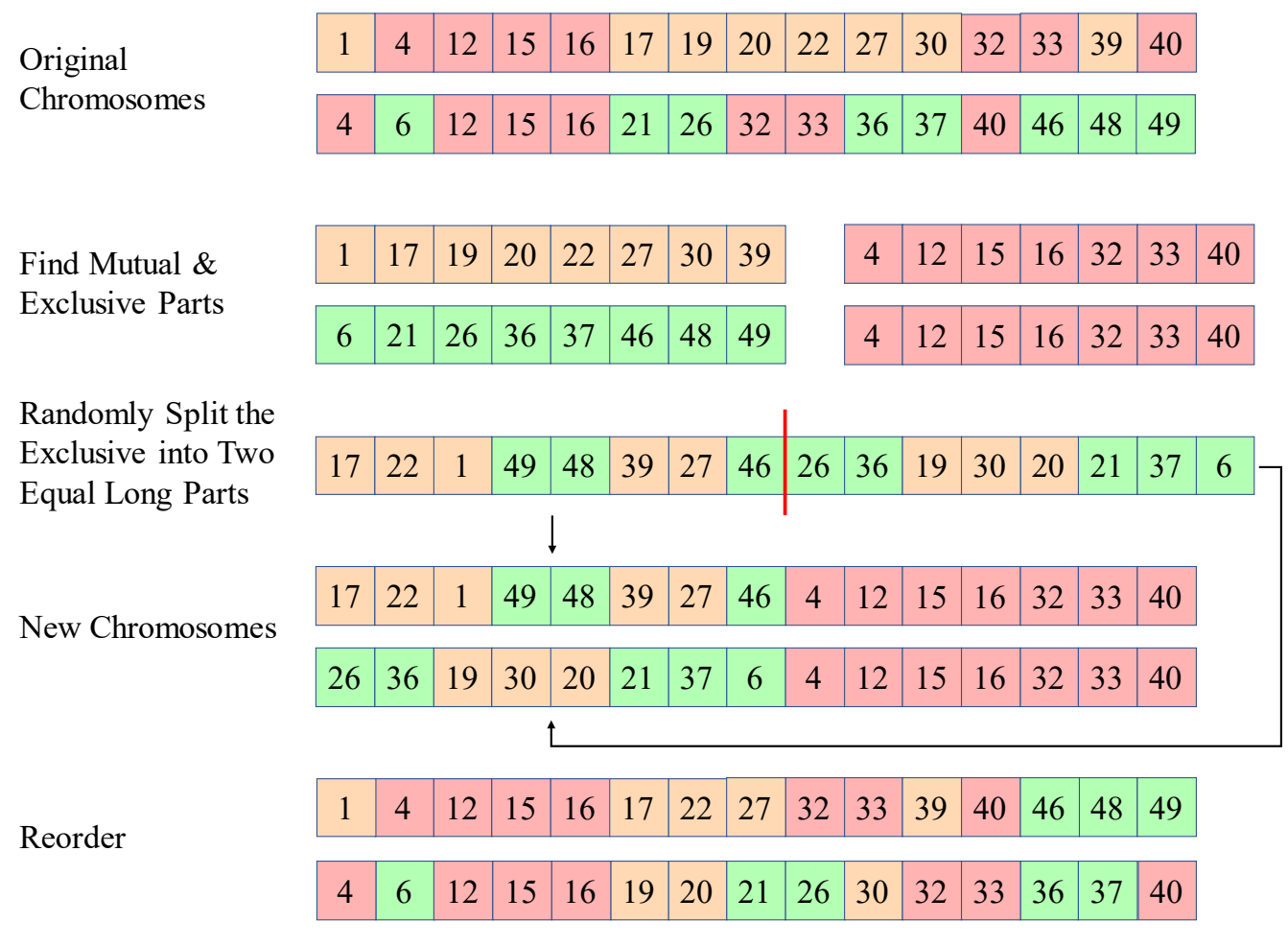

Fig. 3. Process diagram of crossover.

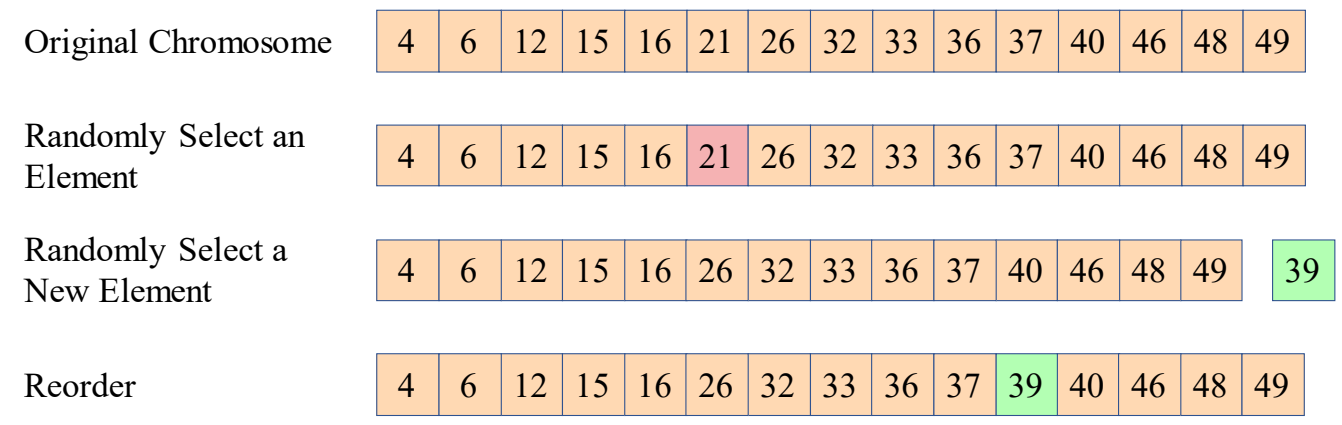

Fig. 4. Process diagram of mutation.

The number of times that a branch is included by a route is recorded and projected to a corresponding color in the "jet" colorbar. With red representing the highest intensity of traffic flow which is about 900 passings and blue the lowest approximately 250, Fig. 7 demonstrates the distribution of traveling demand intensity for each branch.

### B. Results and Analysis

We select a population of size 1500 and set the crossover and mutation probabilities as 0.3 and 0.2 respectively. The large probability values are due to the extremely small ratio of the population size to the total solution space, which includes about $2.5\times10^{17}$ solutions. The maximum iteration is set as 300.

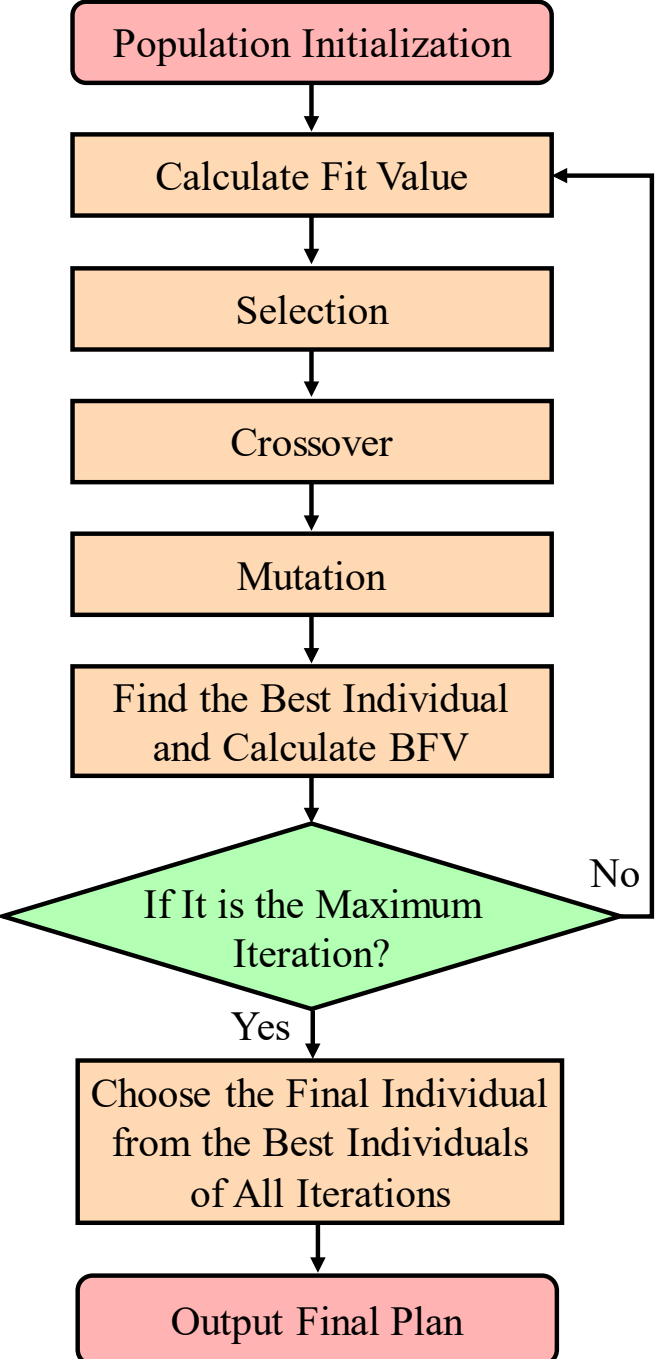

Fig. 5. Flow chart of GA.

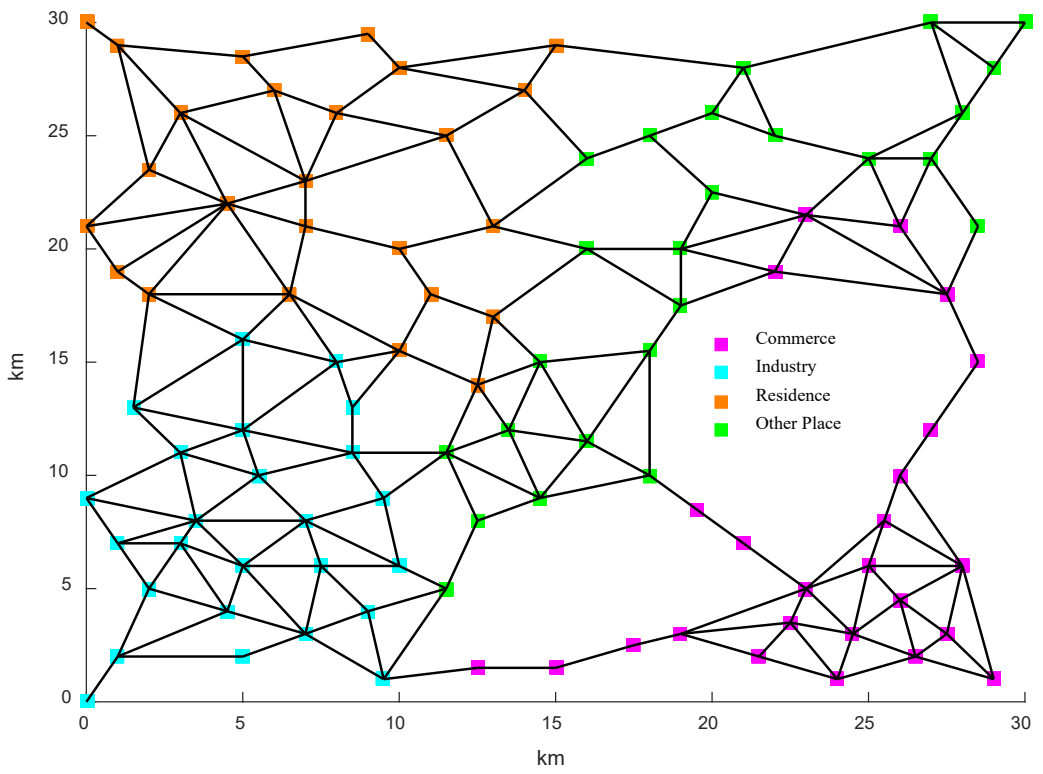

Fig. 6. 100-node 203-branch transportation network.

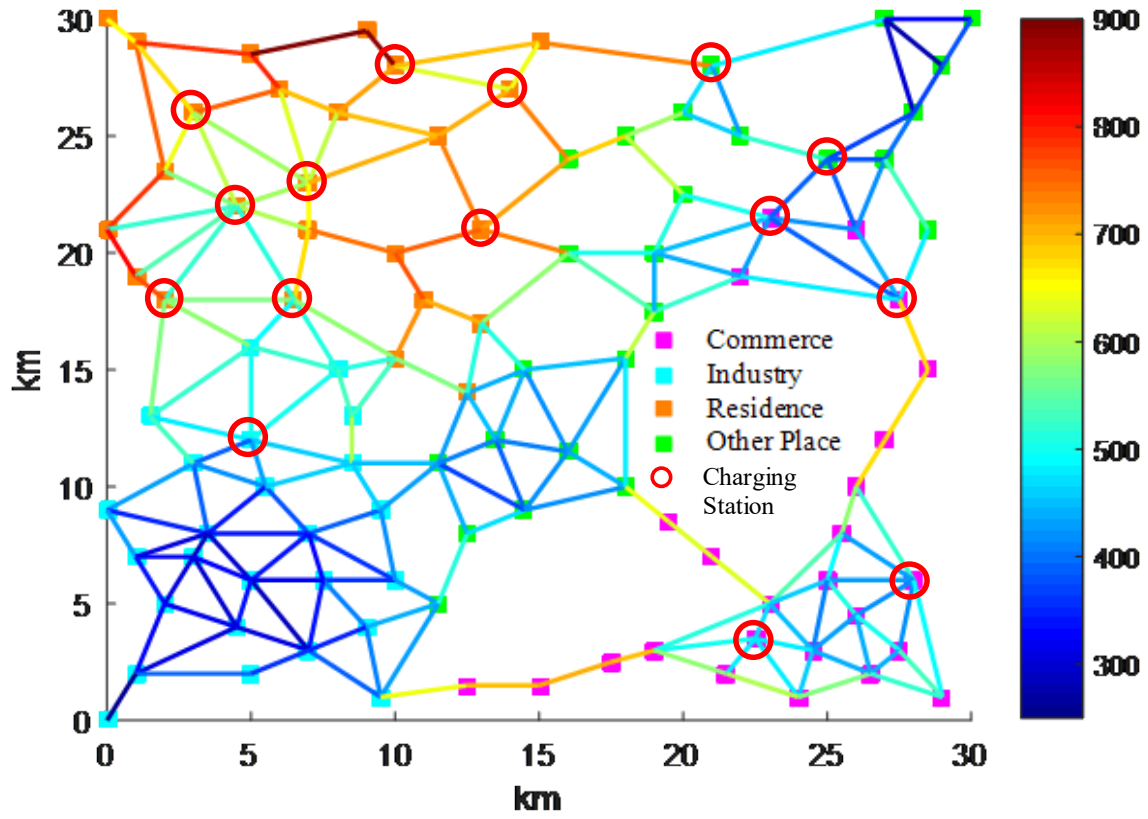

Fig. 7. Branch traffic flow density based on the generated routes.

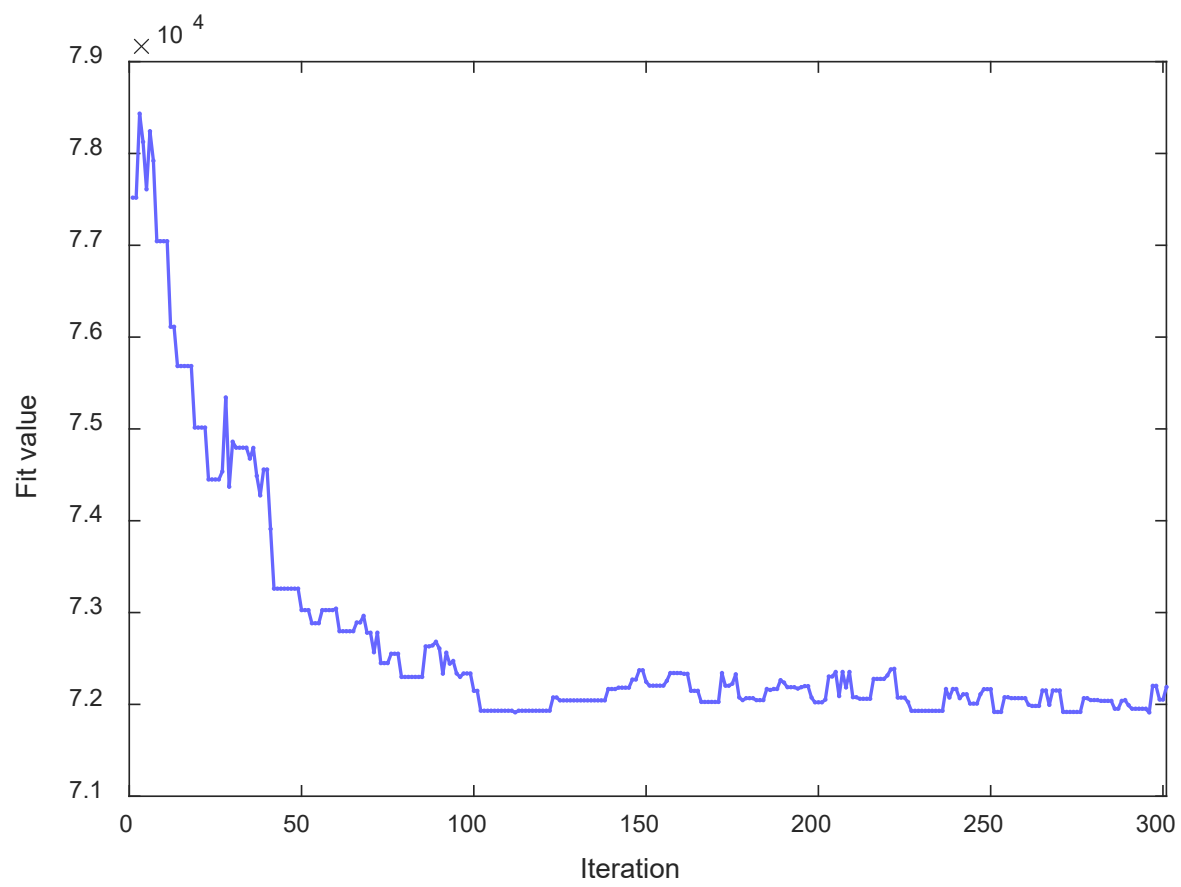

Fig. 8. Best fit values in all iterations.

Fig. 8 shows the best fit value in each iteration, from which we can see that the value declines fast before the 100th iteration and tends to converge after that. The best fit value generates after the 250th iteration and the corresponding station deployment plan is signified by red circles in Fig. 7. The big solution space is the main factor of the slow convergence rate.

From the final station sitting plan, it can be observed that most stations are located at the nodes with high traffic flow or serving as a hub of a small region. For example, two stations are set in the bottom right corner of the network because this area is far from the main part.

The code of our method is edited and run in MATLAB 2018 environment on an iMac with an Intel Core i5 processor and 8 GB random-access memory. For our settings, each iteration costs less than 1 minute and the entire process takes 4 to 5 hours.

## V. Conslusions

In this work, we first propose a methodology of describing the EV detour-to-recharge behavior where a new indicator is created to evaluate EV charging station planning. Second, we come up with an EV charging station planning method using this indicator based on GA. Finally, an assumed transportation network is formulated similar to that of a full functional city, so that the result can be close to that of a real case as possible. The final plan shows that stations are preferred to be located in areas with high traffic flow and at distant places. In future work, the computation efficiency and larger network optimization will be put on the agenda.

## Appendix: Route Generation

In this subsection, the generation of routes in a transportation network is introduced. Before generating a route, an origin area, where the origin node locates in, and a destination area, where the destination node locates in, should be predetermined. We generate the nodes of a route from the origin to the destination through adding next suitable node to a route. First, a node is randomly selected from the origin area as the first node of a route. The last node in a route is regarded as the current node (at first, the randomly selected node is set as the current node) to generate the next node. Before generating the next node of a route, an examination is conducted and show whether this route has met requirements. Three requirements are set: 1) if the last node has been in the destination area; 2) if the length of the route is between the maximum and minimum boundaries; 3) if the route is different from any already generated route. Only when answers of the above three requirements are all "Yes", the route is labeled a valid one. For a valid route, it will be put into the route pool and there is a given probability for it to continue generation of new nodes (note that if a new node is added to a valid route, the route may become invalid again); otherwise, if the route's length is larger than the maximum boundary, the generated will be terminated, and else, it is mandatory to generate new nodes. While generating the next node for a route, based on the current node and other generated nodes, we extract the eligible nodes set, in which any node can be the next node, from the whole nodes in the network according to the following three criteria: 1) the node is connected to the current node by a branch; 2) the node lies within the a special area (explained in the next paragraph) defined by the current node and the destination area; 3) the node has not been visited before, i.e., the node is different from any node which has been in the route, that is, a route including a ring is not considered. All nodes satisfy these three criteria will be put into the eligible nodes set. The final next node will be selected randomly from this set. If the set is empty, this generation will be stopped. The Flow chart of the generation is shown in Fig. A1.

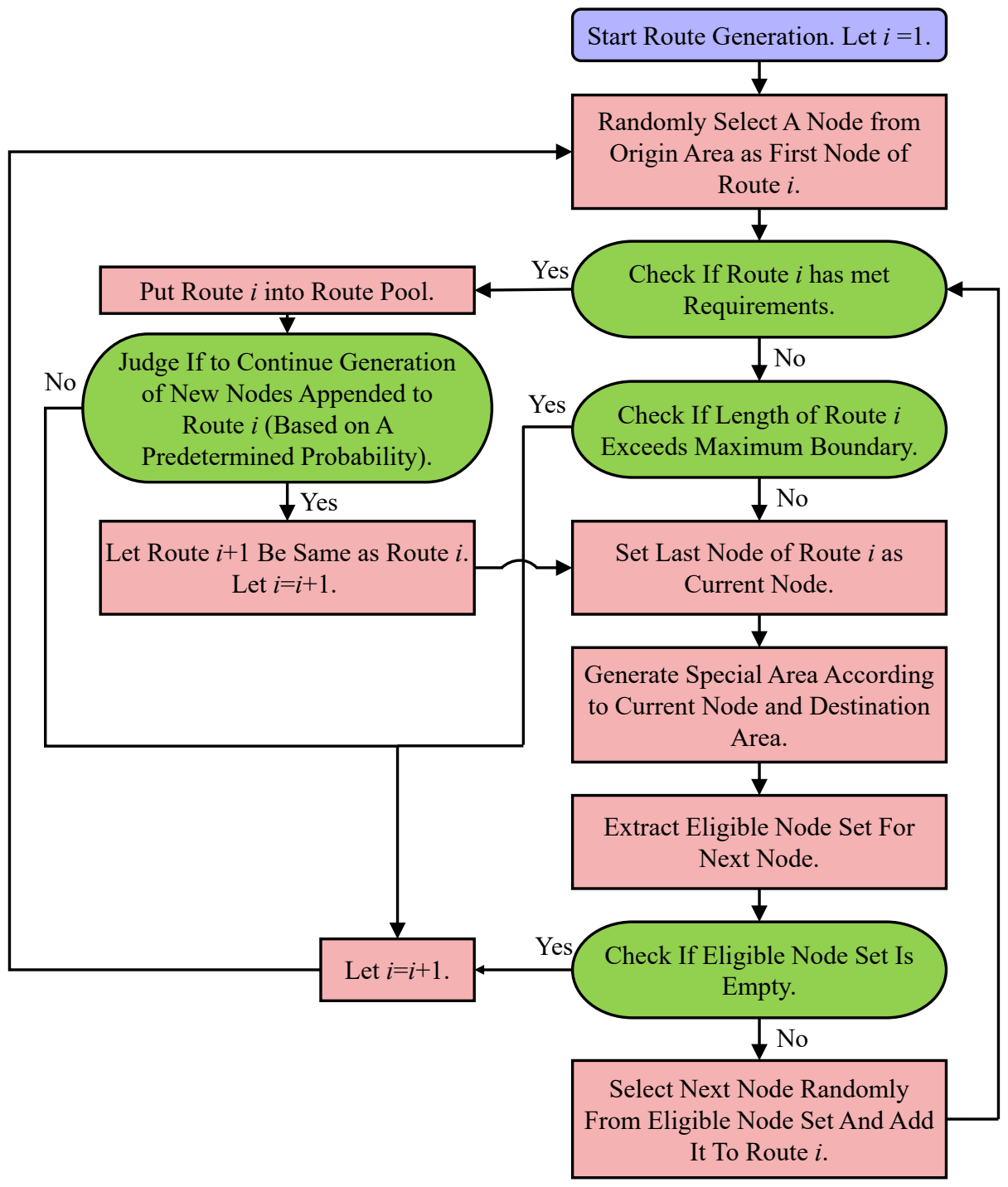

Fig. A1. Flow chart of route generation.

The special area is determined by the locations of the current node and the destination area. First, we identify the circumscribed rectangle of the destination area. Then we divide the surrounding area of the destination area into 8 parts, as Fig. A2 shows. When the current node lies in one of the 4 corner surrounding areas, i.e., areas I, III, V, and VII, connect the current node and three vertexes of the circumscribed rectangle which are the farthest from the current node to form a quadrilateral as the special area. As Fig. A2 shows, we connect the red current node, D1, D3 and D4 (the red quadrilaterals in Fig. A2). When the current node lies in other four surrounding areas, i.e., areas II, IV, VI, and VIII, connect the current node and four vertexes of the circumscribed rectangle to form a pentagon as the special area (the blue pentagon in Fig. A2).

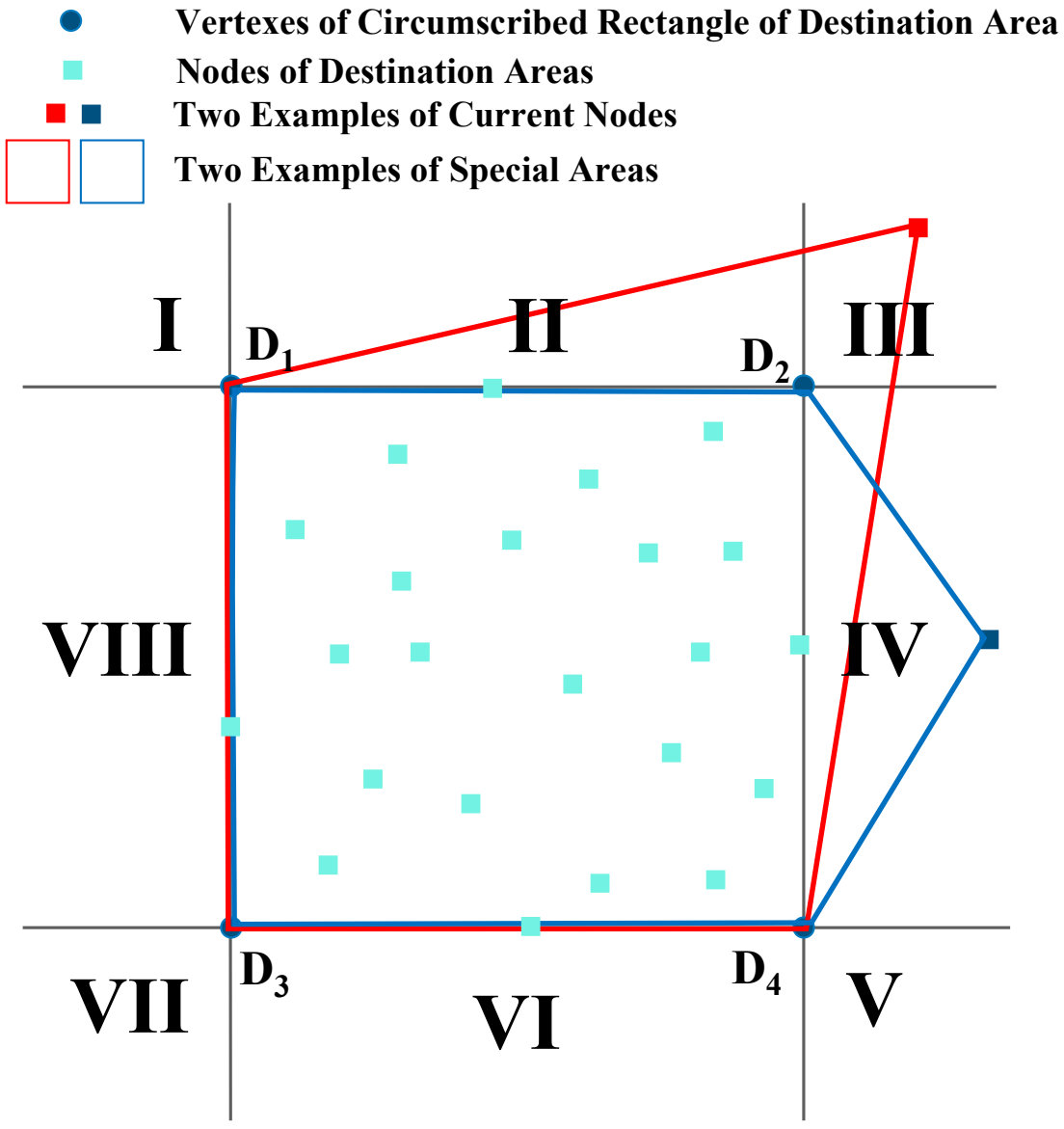

Fig. A2. Diagram of special areas (current nodes outside the circumscribed rectangle).

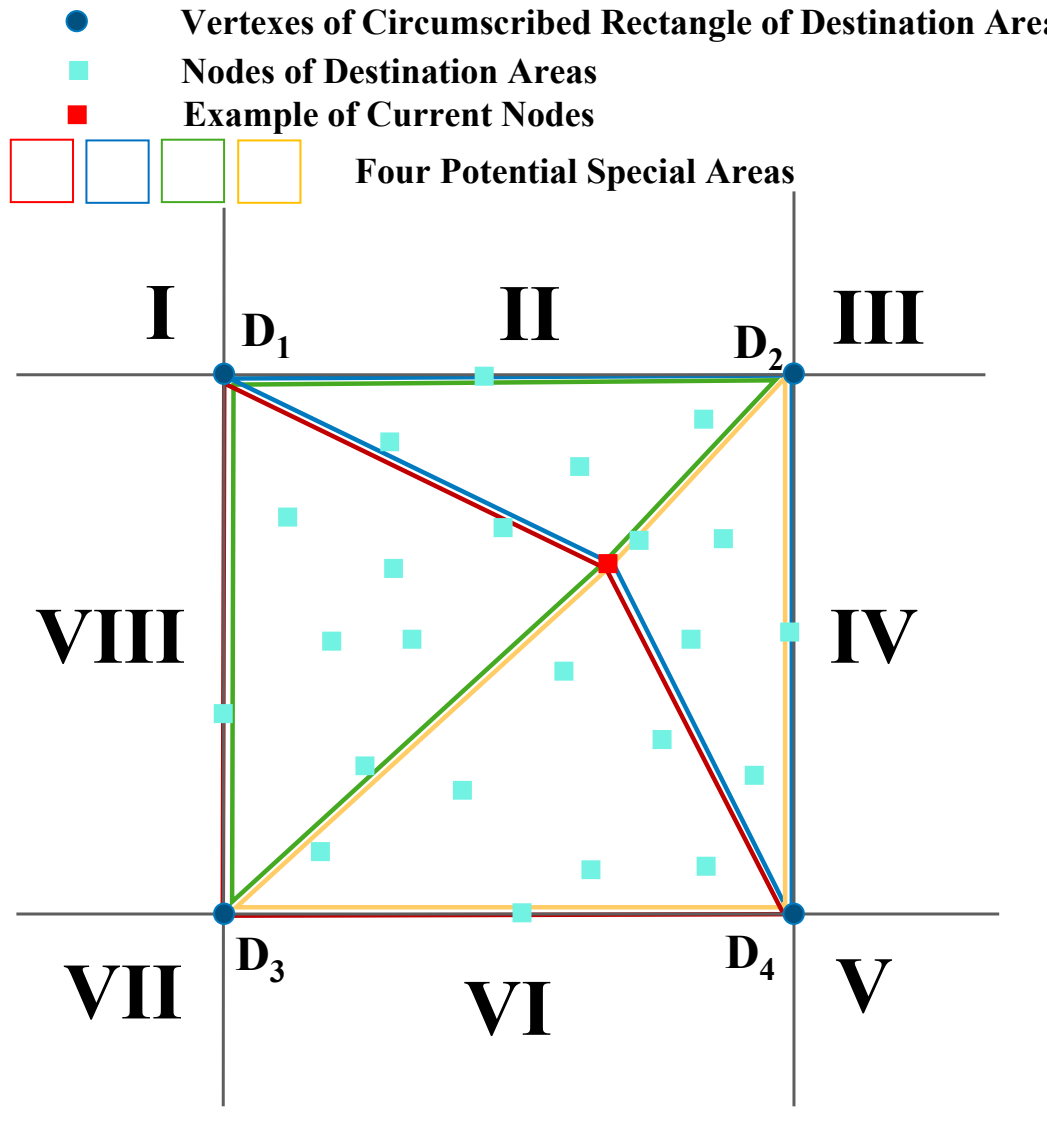

Fig. A3. Diagram of special area candidates (current node inside the circumscribed rectangle).

When the current node lies in the circumscribed rectangle, we connect the current node with three of four vertexes of destination node, and totally form four quadrilaterals. Fig. A3 shows the four quadrilaterals: connecting the current node, D1, D2 and D3 (the green quadrilaterals in Fig. A3); connecting the current node, D1, D2 and D4 (the blue quadrilaterals in Fig. A3); connecting the current node, D1, D3 and D4 (the red quadrilaterals in Fig. A3); connecting the current node, D2, D3 and D4 (the orange quadrilaterals in Fig. A3). Then we extract the quadrilaterals which contains over 40% of the destination area nodes (at least two quadrilaterals) and randomly choose one as the special area.

## REFERENCES

[1] T. Semih and S. Seyhan, "A multi-criteria factor evaluation model for gas station site selection", *Journal of Global Management*, vol 2, no. 1, 2011.

[2] M. Aslani and A. A. Alesheikh, "Site selection for small gas stations using GIS", *Scientific Research and Essays*, vol. 6, no. 15, pp. 1361-3171, 2011.

[3] T. Sweda and D. Klabjan, "An agent-based decision support system for electric vehicle charging infrastructure deployment," in *Proc. IEEE Vehicle Power and Propulsion Conf.*, 2011.

[4] W. Yao et al. "A multi-objective collaborative planning strategy for integrated power distribution and electric vehicle charging systems," *IEEE Trans. on Power Sys.*, vol. 29, no. 4, pp. 1811–1821, 2014.

[5] A. Ip, S. Fong, and E. Liu, "Optimization for allocating BEV recharging stations in urban areas by using hierarchical clustering," in *Proc. 6th Int. Conf. Advanced Inf. Management and Service (IMS)*, Seoul, Korea, 2010.

[6] J. Dong, C. Liu, and Z. Lin, "Charging infrastructure planning for promoting battery electric vehicles: an activity-based approach using multiday travel data," *Transportation Research Part C: Emerging Technologies*, vol. 38, pp. 44-55, 2013.

[7] S. Payam, R. Abbas, and K. Hosein, "Optimal fast charging station placing and sizing," *Applied Energy*, vol. 125, pp. 289-299, 2014

[8] Y. Xiong et al. "Optimal electric vehicle charging station placement," in *Proc. Twenty-Fourth International Joint Conference on Artificial Intelligence*, 2015.

[9] Y. Li, P. Zhang, and Y. Wu, "Public recharging infrastructure location strategy for promoting electric vehicles: A bi-level programming approach," *Journal of Cleaner Production*, vol. 172, pp. 2720-2734, 2018.

[10] S. Bouguerra. and S. B. Layeb, "Determining optimal deployment of electric vehicles charging stations: Case of Tunis City, Tunisia," *Case Studies on Transport Policy*, vol. 7, no. 3, pp. 628-642, 2019.

[11] H. Chen, Y. Jia, Z. Hu, G. Wu, and Z.-J. M. Shen. "Data-driven planning of plug-in hybrid electric taxi charging stations in urban environments: A case in the central area of Beijing," In *Proc. 2017 IEEE PES Innovative Smart Grid Technologies Conference Europe (ISGT-Europe)*, 2017.